\documentclass[aps,prd,floatfix,nofootinbib,superscriptaddress,twocolumn,eqsecnum,tightenlines]{revtex4-1}
\usepackage[mathscr]{euscript}
\usepackage{amsmath,amsfonts,amssymb}
\usepackage{bm}
\usepackage{graphicx}% Include figure files
\usepackage[pdftex]{color}
\usepackage[sort&compress]{natbib}
\usepackage[colorlinks=true,linkcolor=blue,filecolor=blue,urlcolor=blue,citecolor=blue,pdftex,plainpages=false]{hyperref}
\usepackage{multirow}

\newcommand{\sea}{\text{sea}}

\newcommand{\Xsl}[1]{\raise.15ex\hbox{/}\kern-.57em #1}

\DeclareMathOperator{\Tr}{Tr} % AMS

\newcommand{\mbar}{\overline{m}}
\newcommand{\asPI}{\frac{\alpha_s}{\pi}}

\newcommand{\lat}{\text{lat}}

\newcommand{\PBP}{{\langle \overline{\psi}\psi \rangle}}
\newcommand{\SBS}{{\langle \overline{s}s \rangle}}

\newcommand{\GeV}{\text{GeV}}
\newcommand{\MeV}{\text{MeV}}

\newcommand{\strike}[1]{}

\begin{document}

\title{Determination of the quark condensate from heavy-light current-current correlators in full lattice QCD}

\author{C.~T.~H.~Davies}
\email[]{christine.davies@glasgow.ac.uk}
\affiliation{SUPA, School of Physics and Astronomy, University of Glasgow, Glasgow, G12 8QQ, UK}
\author{K.~Hornbostel}
\email[]{deceased}
\affiliation{Southern Methodist University, Dallas, Texas 75275, USA}
\author{J.~Komijani}
\email[]{javad.komijani@glasgow.ac.uk}
\affiliation{SUPA, School of Physics and Astronomy, University of Glasgow, Glasgow, G12 8QQ, UK}
\author{J.~Koponen}
\affiliation{INFN, Sezione di Tor Vergata, Dipartimento di Fisica, Universit\`{a} di Roma Tor Vergata, Via della Ricerca Scientifica 1, I-00133 Roma, Italy}
\author{G.~P.~Lepage}
\affiliation{Laboratory for Elementary-Particle Physics, Cornell University, Ithaca, New York 14853, USA}
\author{A.~T.~Lytle}
\affiliation{SUPA, School of Physics and Astronomy, University of Glasgow, Glasgow, G12 8QQ, UK}
\affiliation{INFN, Sezione di Tor Vergata, Dipartimento di Fisica, Universit\`{a} di Roma Tor Vergata, Via della Ricerca Scientifica 1, I-00133 Roma, Italy}
\author{C.~McNeile}
\affiliation{Centre for Mathematical Sciences, University of Plymouth, PL4 8AA, UK}
\collaboration{HPQCD collaboration}
\homepage{http://www.physics.gla.ac.uk/HPQCD}
\noaffiliation

\date{\today}

\begin{abstract}
We derive the Operator Product Expansion whose vacuum expectation 
value gives the time-moments of the 
pseudoscalar heavy-light current-current correlator 
up to and including terms in $\alpha_s^2$ multiplying 
$\langle\overline{\psi}\psi\rangle/M^3$ 
and terms in $\alpha_s$ multiplying $\langle \alpha_s G^2 \rangle/M^4$, where 
$M$ is the heavy-quark mass. 
Using lattice QCD results for heavy-strange correlators 
obtained for a variety of heavy quark masses 
on gluon field configurations including $u$, $d$ and $s$ quarks 
in the sea at three values of the lattice spacing, we are able to show that 
the contribution of the strange-quark condensate to the time-moments is very substantial.  
We use our lattice QCD time-moments and the OPE to determine a value for the condensate,
 fitting the 4th, 6th, 8th and 10th time-moments
simultaneously. Our result, 
$\langle \overline{s}s \rangle^{\overline{\text{MS}}}(2 \text{GeV}) =  -(296(11) \,\mathrm{MeV})^3$,
agrees well with HPQCD's earlier, more direct, lattice QCD determination~\cite{McNeile:2012xh}.
As well as confirming that the $s$ quark condensate is close in value to the 
light quark condensate, this demonstrates clearly the consistency of the Operator Product Expansion 
for fully nonperturbative calculations of 
matrix elements of short-distance operators in lattice QCD.
\end{abstract}

\maketitle

\section{Introduction}
\label{sec:intro}

The physics of the strong interaction clearly exhibits the features of 
spontaneously broken chiral symmetry induced by the condensation 
of quark-antiquark pairs in the vacuum.
The fact that the vacuum expectation value of $\overline{\psi}\psi$ is 
non-zero is readily demonstrated in the fully nonperturbative approach 
to QCD provided  by lattice QCD~\cite{DeGrand:2006zz}. Calculating 
an accurate value for $\langle 0 | \overline{\psi}\psi | 0 \rangle$ is, however, 
far from simple as we will discuss further below. 

Each of the light quark flavours, $u$, $d$ and $s$, gives a condensate, 
which could differ in value because of the different quark masses.
The value of the condensate at zero quark mass (the chiral condensate) is 
an important parameter of low energy QCD~\cite{Reinders:1984sr}. 
Coefficients corresponding 
to derivatives of the chiral condensate with respect to quark mass then appear 
in higher order terms in the chiral expansion. Alternatively, and 
more simply, one can determine values for the condensates of specific quarks 
at their physical 
quark masses and these are then important input for analyses using QCD sum rules. 
For example, the values of both the $s$ quark condensate and
the light ($u/d$) quark 
condensate are key ingredients in the determination of $|V_{us}|$ from 
hadronic $\tau$ decay~\cite{Hudspith:2017vew}.  
These condensates can be determined most accurately by lattice QCD
calculations~\cite{McNeile:2012xh} and we give a new method for doing this here. 
We will apply the method to determine the strange quark condensate, 
although the approach could also 
be used for the light quark case. 

One way in which the nonzero value of the quark condensate 
feeds into strong interaction physics is through its appearance 
in sub-leading (`nonperturbative') terms in the Operator Product Expansion (OPE) 
of short-distance quantities~\cite{Wilson:1969zs, Shifman:1978bx, Reinders:1984sr, Shifman:1998rb}. 
The OPE means that care must be taken to understand condensate 
contributions when determining such quantities in a fully nonperturbative 
calculation in lattice QCD. It also means that values for condensates can 
be determined from such calculations provided that sufficient accuracy is 
available to separate the different orders in the OPE. 
The determination of the same condensates from multiple short-distance 
quantities provides a strong test of the OPE approach. 

The quark condensate is the lowest 
dimensional gauge-invariant one, so might be expected to 
dominate the subleading terms of calculations of 
gauge-invariant quantities\footnote{
The Landau gauge gluon condensate 
has lower dimension and must be taken into account in lattice 
QCD calculations of gauge-noninvariant quantities~\cite{Chetyrkin:2009kh, Blossier:2010ky, Burger:2012ti, Lytle:2018evc}.}. 
In fact quark condensate contributions are often rather small 
because $\overline{\psi}\psi$ will typically appear in a chirally-symmetric 
combination with the quark mass, $m$~\cite{Reinders:1984sr}. 
This then requires an additional inverse power of the high momentum parameter, $Q$, 
which defines the short-distance.  
For light quarks with $m << \Lambda_{QCD}$ the  
$m/Q$ factor provides significant suppression. When a valence light 
quark is combined with a heavy quark, however, the spin of the heavy quark can 
be flipped with very little penalty, to compensate the light quark spin flip
in $\overline{\psi}\psi$ and no factor of $m$ appears. 
This implies that short-distance quantities derived from 
heavy-light correlation functions 
should provide a good signal for the light quark condensate, if an 
accurate OPE can be constructed.

Here we demonstrate such an analysis using time-moments of 
pseudoscalar heavy-strange current-current correlators 
and an $\alpha_s^2$-accurate OPE in the $\overline{\text{MS}}$ scheme that extends 
through NNLO in inverse powers 
of the heavy-quark mass (the short-distance scale here). 
The moments can be calculated very precisely in lattice 
QCD for a range of heavy quark masses up to the $b$. 
We use moments whose values are independent 
of the lattice spacing, up to discretisation effects that can be removed.  
The observed dependence on the
heavy quark mass, matched to that expected from the 
OPE then allows us to obtain 
the strange quark condensate in the $\overline{\text{MS}}$ scheme. 
We fit the four lowest moments simultaneously for our final answer.  

We can compare this method to earlier results 
from a more direct lattice QCD calculation~\cite{McNeile:2012xh}. 
The earlier calculation determined 
the expectation value of the trace of the quark propagator in lattice QCD. 
Naively this quantity would appear to be the quark condensate, 
but the mixing between the $\overline{\psi}\psi$ operator and the identity 
means that an OPE had also to be developed for this case. 
This was used to relate the lattice result to the strange quark 
condensate in the continuum 
in the $\overline{\text{MS}}$ scheme. The agreement between our new result 
and the earlier one
is then a test of the OPE approach.  

The paper is laid out as follows: Section~\ref{sec:theory} gives the theoretical 
background and the construction of the OPE, Section~\ref{sec:latt} describes the lattice calculation and 
Section~\ref{sec:analysis} the analysis that combines the two. Finally Section~\ref{sec:conclusions}
compares our result for the condensate to earlier values and gives 
our conclusions.

\section{Theoretical background}
\label{sec:theory}

The time-moments of current-current correlators at zero 
spatial momentum can be related to $q^2$-derivative moments 
of the corresponding polarisation functions. 
They are physical quantities, i.e. they do not depend 
on the ultraviolet regulator, 
and their values can be determined from the 
continuum and physical $u/d$ quark mass limit 
of lattice QCD calculations. 
The HPQCD collaboration has pioneered the lattice calculation of 
such moments and their comparison with values, in 
the vector current case, derived from 
experimental results for
the cross-section for $e^+e^- \rightarrow \text{hadrons}$ 
via a virtual photon as a function of centre-of-mass energy~\cite{Donald:2012ga, Chakraborty:2014mwa, Chakraborty:2016mwy}. 
See also Refs.~\cite{Nakayama:2016atf, Borsanyi:2016lpl}. 

When the currents contain heavy quark fields, the low 
moments are perturbative, with the scale of $\alpha_s$ being set by 
the heavy quark mass. Each moment is given by a power series in 
$\alpha_s$, 
multiplying an inverse power (depending on the moment number) of the quark mass. 
For heavyonium currents (made of a heavy quark and antiquark)
the perturbative series is known to high order (through 
$\alpha_s^3$)~\cite{Chetyrkin:2006xg, Boughezal:2006px, Maier:2008he, Kiyo:2009gb, Maier:2009fz}. This then provides an accurate method 
for the determination of heavy quark masses by comparing 
the perturbation theory to nonperturbative 
results for the moments either from experiment for $e^+e^- \rightarrow \text{hadrons}$ 
(see, for example, Ref.~\cite{Kuhn:2007vp}) or from lattice 
QCD~\cite{Allison:2008xk, McNeile:2010ji, Chakraborty:2014aca, Nakayama:2016atf, Maezawa:2016vgv}.  
The results using experimental information are necessarily 
restricted to the vector (electromagnetic) current but
lattice QCD results are available for currents with other spin-parity. 
The pseudoscalar current-current correlator is particularly 
useful in that case since, in lattice QCD formalisms 
with sufficient chiral symmetry, the quark mass times current 
is absolutely normalised as in continuum QCD, removing systematic uncertainties 
from current normalisation. 

For the pseudoscalar case the moments are:
\begin{equation}
\label{eq:mdef}
\mathcal{M}_n = M^2 \int d\mathbf{x}\,dt\,t^n\,
\langle 0 | J_5(\mathbf{x},t)J_5(0) | 0 \rangle .
\end{equation}
Here $M$ is the heavy quark mass, $J_5 = \overline{\psi}_h\gamma_5\psi_h$ with 
$\psi_h$ the heavy quark field and the integral over $\mathbf{x}$ projects 
onto zero spatial momentum. 
The first four $q^2$-derivative moments,  1, 2, 3 and 4 correspond to $n=$ 4, 6, 8 and 10.  
In perturbation theory $\mathcal{M}_n$ is given by 
\begin{equation}
\label{eq:Mseries}
\mathcal{M}_n = \frac{g_n(\alpha_s(\mu),\mu/M)}{(M(\mu))^{n-4}} ,
\end{equation}
where $g_n$ is a power series in $\alpha_s$. 

An important advantage of the heavyonium current-current correlator 
technique for determination of quark masses is the 
insensitivity to nonperturbative contributions coming from condensates. 
The radiative corrections to the 
heavy-quark vacuum polarisation and its derivatives close to 
$q^2$ = 0~\cite{Shifman:1978bx, Reinders:1984sr} are 
sensitive to phase-space regions of quark momentum, $p$, and 
gluon momentum $k$ where $p^2 \rightarrow 0$ and $k^2 \rightarrow 0$. 
Since heavy quarks are still highly virtual as $p^2 \rightarrow 0$, 
it is only the $k^2 \rightarrow 0$ region that is sensitive to long-distance 
physics.  
This generates a contribution to heavyonium current-current 
correlator moments given by the gluon condensate divided by 
four powers of the heavy quark mass. 
For the low moments used for the determination 
of the quark mass, the effect of this term on the moments 
is below 0.05\% even when the 
heavy quark is a charm quark, and it has negligible impact~\cite{Chakraborty:2014aca}. 

In this work we study time-moments of heavy-light current-current 
correlators. We will see that nonperturbative condensate contributions, 
in this case coming from the light quark condensate, have about 
one hundred times more impact. This is largely because the leading quark condensate 
term appears with only 
three inverse powers of the heavy quark mass 
and no powers of $\alpha_s/\pi$ (and, as discussed earlier, no powers of the light quark 
mass). Heavy-light moments 
then in fact present an opportunity for a determination 
of the light quark condensate. To 
do this we first need to determine an accurate OPE
to compare to our lattice QCD results. This we do in the next section.  

\subsection{The OPE for heavy-light current-current corrrelators}
\label{subsec:OPE}

Appendix B of Ref.~\cite{McNeile:2012xh} discusses what is needed to derive the OPE for heavy-light 
current-current correlators.
For ease of reference, we summarise the pieces that we will use here.   

Consider moments of two pseudoscalar densities, $J_5 = \overline{\psi}_h\gamma_5\psi$, 
composed of a heavy quark (mass~$M$) and a light quark (mass~$m\ll  M$) 
where the heavy quark fields are contracted with each other:
\begin{equation}
\label{eq:momcont}
M^2 \int d\mathbf{x}\,dt\,t^n\,
J_5(\mathbf{x},t)J_5(0) \rightarrow \mathcal{O}^{(n)} ,
\end{equation}
where
\begin{equation}
   \mathcal{O}^{(n)} \equiv \int d\mathbf{x}\,dt\,t^n\,
   \bar{\psi}(\mathbf{x},t)\gamma_5
   \,\frac{M^2}{D\cdot\gamma+M}\,\gamma_5
      \psi(0).
\end{equation}
As discussed above in the heavyonium case, the $M^2$~factor 
makes $\mathcal{O}^{(n)}$ independent of the 
ultraviolet regulator provided $n\!\ge\!4$. This means that lattice and 
continuum calculations should agree in the limit of zero lattice spacing.
Continuum results derived from lattice calculations can then be compared 
to continuum expressions derived from continuum QCD perturbation theory. 

Operator~$\mathcal{O}^{(n)}$ is short-distance, dominated by length scales of 
order~$1/M$, provided the heavy-quark is sufficiently heavy and the 
light quarks have momenta small compared with~$M$. 
Consequently the~OPE implies that~$\mathcal{O}^{(n)}$ can be expressed in terms 
of a set of local operators in an effective theory, with 
cutoff scale~$\Lambda\!<\!M$, and coefficient functions 
that depend only upon physics between scales~$\Lambda$ and~$M$:
\begin{align} \label{ope}
   M^{n-4}\mathcal{O}^{(n)} &= \mathbf{1}^{(\Lambda)}\, 
   c_n(\Lambda/M,\alpha_s,m/M)
    \nonumber\\
   &+ \frac{(\overline{\psi}\psi)^{(\Lambda)}}{M^3}\,
   d_n(\Lambda/M,\alpha_s,m/M) \\
   &+ \frac{(\alpha_s G^2)^{(\Lambda)}}{M^4}\,
   e_n(\Lambda/M,\alpha_s,m/M) \nonumber \\
   &+\ldots.\nonumber
\end{align}
 $\mathbf{1}^{(\Lambda)}$ is the unit operator and note that $\overline{\psi}\psi$ is 
{\it not} normal-ordered~\cite{Chetyrkin:1985kn, Jamin:1992se}. 
Working in the continuum we take the effective theory on the right-hand side of Eq.~(\ref{ope}) 
to be QCD with an~$\overline{\mathrm{MS}}$ regulator and~$\Lambda\!=\!\mu$.
Then we can express the right-hand side in terms of masses and couplings at 
scale $\mu$, often chosen so that $\mu\!=\!M(\mu)$.

The coefficient functions~$c_n$,~$d_n$ and $e_n$ are perturbative when~$M$ is large, 
and analytic in~$\alpha_s$ and~$m/M$.
They can be determined by perturbative matching of the matrix elements of the 
left- and right-hand sides of Eq.~(\ref{ope}) between a given set of states. 
What we expect to see is that the matrix element of the left-hand side exhibits some 
infrared sensitivity that can be recognised as being part of the perturbative expression 
for one of the condensates on the right-hand side, requiring a particular value for 
the coefficient multiplying that condensate.  

Here we can use vacuum matrix elements for the perturbative matching 
because the perturbation theory that we need 
in that case has already been done. 
We take the perturbative expressions for the vacuum matrix element of the left-hand-side 
of Eq.~(\ref{ope}) as a function of 
$x \equiv m/M$ from Refs.~\cite{Hoff:2011ge, Grigo:2012ji}. We can identify 
the condensate contributions 
from the non-analytic pieces  
that contain $\ln x$ multiplied by powers of $x$~\cite{Tkachov:1999nk}. This can be done 
straightforwardly and unambiguously through multiple orders in 
$\alpha_s$ and $x$ because sufficiently accurate 
perturbative expressions for the light quark condensate are 
also known~\cite{Braaten:1991qm, Kneur:2015dda}. The gluon condensate does not appear  
until $\alpha_s^2 x^4\ln x$ and so there are no issues with the separation 
of terms that can be identified 
with each condensate.  

The starting point for the OPE derivation is  (from Eq.~(\ref{ope}))
\begin{eqnarray}
\label{eq:vevope}
\langle \frac{M^{n-4}}{n!}\mathcal{O}^{(n)} \rangle &&\equiv \frac{M^{n-4}}{n!} \mathcal{M}_n \\
  && = g_n(x,\alpha_s(\mu), \mu/M) \nonumber \\ 
&& = c_n + d_n \frac{\langle \overline{\psi}\psi \rangle}{M^3}
  + e_n \frac{\langle \alpha_s G^2 \rangle }{M^4} + \ldots , \nonumber 
\end{eqnarray}
where we take out a factor of $n!$ for future convenience. 
The perturbative expansion of the left-hand side is given by
the expressions in Refs.~\cite{Hoff:2011ge, Grigo:2012ji} 
for ${3/(4\pi)^2 \times}\overline{C}_k^p$ for $2k+2=n$. 

The perturbative expansion through $\alpha^2_s(\mu)$ for the vacuum matrix element 
of $\overline{\psi}\psi$ for a quark of mass $m$ is given in Ref.~\cite{Kneur:2015dda}:
\begin{align}
\label{eq:PBP:pert}
  \PBP (\mu) &= \frac{m^3}{4\pi^2} \bigg[ 
      ( 3 - 6 L_m) + \\
      &\hspace{-10pt}\asPI \left(10 - 20 L_m + 24 L_m^2\right) + \nonumber \\ 
      & \hspace{-18pt}\left(\asPI\right)^2 \bigg(98.739 + 0.803 n_f 
       + (-178.775 + \frac{39}{2} n_f) L_m  \nonumber \\
      &+ (188 - \frac{20}{3} n_f ) L_m^2 + (-108 + \frac{8}{3} n_f) L_m^3 \bigg)
      \bigg] , \nonumber
\end{align}
where $L_m = \ln(m/\mu)$ and $n_f=n_l+n_m$, with $n_l$ massless quarks and $n_m$ quarks with
mass $m \equiv m(\mu)$ in the sea.

This allows us to demonstrate simply how the OPE is derived using part of the perturbative 
expansion for one of the moments. We consider the zeroth and first order in $\alpha_s$ 
expansions for the first ($n$=4) moment for the heavy-light current-current correlator.
The expressions given in Ref.~\cite{Hoff:2011ge} as a function of $x=m/M$ can be 
expanded out to fourth order in $x$ to give, through $\mathcal{O}(\alpha_s)$ for the 
fourth moment:
\begin{align}
\label{eq:hoff}
\frac{(4\pi)^2}{4!\,3}\mathcal{M}_4 &= 0.667 + 0.667x - 2x^2  \nonumber \\
& + x^3(8\ln x + 9.333) - x^4(16\ln x + 16.667)  \nonumber \\ 
& + \frac{\alpha_s}{\pi} \bigg[ 1.573 + 5.240x -12.240 x^2 \nonumber \\
& - x^3(32\ln^2 x + 5.333\ln x - 32.960) \nonumber \\ 
& + x^4(64\ln^2 x + 8 \ln x - 57.200) \bigg] .
\end{align}
The terms with $\ln x$ indicate infrared sensitivity of the series, which will be made 
explicit when these terms are replaced by the quark condensate. In Eq.~(\ref{eq:PBP:pert}) we 
can replace $L_m$ with $\ln x + \ln(M/\mu)$ to expose terms in $\ln x$. This allows us 
to substitute $\PBP$ for the leading $x^3\ln x$ term on the second line of Eq.~(\ref{eq:hoff}).   
In doing that we find that the $x^3\ln^2 x$ at $\mathcal{O}(\alpha_s)$ 
is automatically absorbed into $\PBP$. Likewise the $x^4 \ln x$ term at zeroth order can 
be replaced by a term proportional to $x \PBP$ and this automatically absorbs the 
term at $\mathcal{O}(\alpha_s)$ of the form $x^4 \ln^2 x$. Notice that none of this 
could be done if the perturbative calculation had set the light quark mass (and hence $x$) 
to zero.  

Our series for the fourth moment through $x^4$ then becomes
\begin{eqnarray}
\label{eq:opeseries}
\frac{\mathcal{M}_4}{4!} &=& 0.0127 + 0.0127 x -0.0380 x^2 + 0.253 x^3 -0.469x^4 \nonumber \\
&&\hspace{-20pt}+ \frac{\alpha_s}{\pi} [ 0.0299 + 0.0996 x -0.233 x^2 + 0.576 x^3 -1.011 x^4 ] \nonumber \\
&&+ \frac{\PBP}{M^3}[-1 + 2x + \frac{\alpha_s}{\pi}(4-7.667 x)] .
\end{eqnarray}
The expansion clearly has the form expected%
\footnote{The $(-1+2x)$ multiplying the quark condensate at $\mathcal{O}(\alpha^0_s)$
  was derived earlier in~\cite{Reinders:1980wk, Reinders:1984sr}.}
in Eq.~(\ref{eq:vevope})
where the short-distance coefficients $c_n$, $d_n$, $e_n$ and their higher order counterparts
are power series in $\alpha_s$ with only polynomial dependence on $x$.
We identify the series for $d_4$ in the square brackets of the third line. 
Notice that the leading term of $d_4$ is independent of $x$, i.e. $\PBP$ appears without 
powers of the quark mass. 
It is also clear that the size of the coefficients appearing in $d_4$, compared to those in $c_4$, 
emphasise the condensate contribution to $\mathcal{M}_4$. This is what allows us 
to `see' the effect of the condensate in the heavy-light moments so clearly. 

The results in Refs.~\cite{Hoff:2011ge} and~\cite{Kneur:2015dda} in fact allow us to 
derive the $c_n$ and $d_n$ series through $\alpha^2_s$ and, respectively, $x^4$ and $x$. 
These series are given in Tables~\ref{tab:c_n:hs} and~\ref{tab:d_n:hs} 
in Appendix~\ref{appendix:coeffs}. We give results for $n=4$, 6, 8 and 10. 
The relative size of the $d_n$ coefficients compared to $c_n$ grows 
with $n$. This is consistent with the expectation that longer-distance nonperturbative 
effects play an increasing r\^{o}le in the higher moments where there is a bigger 
contribution from larger $t$ values. 

The perturbation theory for heavy-light moments has been extended through 
$\alpha_s^3$ in~\cite{Maier:2015jxa} for massless quarks. Although this does not allow us to determine
condensate contributions it does allow the leading $x$-independent term to 
be added to $c_n$ and we will test the impact of that in Section~\ref{sec:analysis}.  

At $\mathcal{O}(\alpha^2_s)$ 
quarks in the sea start to contribute. It is then important to distinguish between 
the number of massless quarks, $n_l$, the number of quarks of mass $m$, $n_m$, 
and the number of heavy quarks, $n_h$, active in the sea. 
Ref.~\cite{Hoff:2011ge} provides a Mathematica script giving 
their calculation at $\alpha^2_s$ with these pieces explicit. 
Although we derive here the OPE for heavy-light correlator moments for a light 
quark of mass $m$ in terms of the condensate for that quark, the massless quarks 
in the sea will also condense. Because their masses have been set to zero in 
the perturbative expansion it is not possible to explicitly extract the 
contribution from their condensate. This is, however, a very small effect because 
these quark condensate contributions must appear multiplied by the light quark mass, 
as a consequence of chiral symmetry. Hence they can be neglected. 
This is discussed 
further in Section~\ref{sec:analysis}.  

The perturbative expansion at $\alpha^2_s$ allows the term multiplying 
the gluon condensate to be determined at leading order, using the perturbative 
expansion for the gluon condensate from Ref.~\cite{Braaten:1991qm}
\begin{align}
\label{eq:gluecondensate}
  \langle \alpha_s G^2 \rangle(\mu)  &= 
    - \frac{n_m m^4}{2\pi} \left(\asPI\right)^2 \left(9 - 16 L_m + 12 L_m^2\right) .
\end{align}
We find $e_n = 1/(12\pi)$ 
in agreement with~\cite{Reinders:1980wk, Reinders:1984sr}. 

The OPE tells us that the behaviour of the heavy-light moments is very different 
from expectations from a naive application of the $x$-dependent perturbative expansion.
We will be able to demonstrate that clearly in Section~\ref{sec:latt} from our 
lattice QCD results. For light quarks of small mass $x$ will be small; it is less than 
0.05 for the range of quark masses we use in our calculation. 
Powers of $x$ and powers of $x$ multiplying $\ln x$ then have relatively small impact. 
The OPE tells us, however, that in a fully nonperturbative scenario such as the real 
world or a lattice QCD calculation, these terms will be replaced by much 
larger terms coming from the light quark condensate. The coefficients with which the 
condensate terms appear, especially the fact that they enter at $\mathcal{O}(\alpha^0_s)$, 
means that the terms are clearly visible in the results for the heavy-light moments.   
The OPE requires that the light quark condensate that enters here is the same 
matrix element as appears in the OPE for other operators and this consistency check 
of the OPE framework is an important one.  Accurate nonperturbative results and 
an OPE that goes well beyond leading order are both required for this. The OPE derived 
here will be used in the next section along with our lattice QCD calculation to 
determine a value for the $s$ quark condensate.

\section{The lattice QCD calculation} 
\label{sec:latt}

In the Section~\ref{sec:theory} we discussed how the OPE leads to time-moments 
of heavy-light current-current correlators containing important terms proportional 
to the light quark condensate. Here we describe how we calculate these time moments 
fully nonperturbatively in a lattice QCD calculation. We will focus on 
heavy-strange correlators and the determination of the $s$ quark condensate, 
because that is computationally simpler than for $u/d$ quarks.
In fact, as we shall see, it is convenient to take a ratio of the heavy-strange 
moments to those of heavy-charm current-current correlators. Both
are calculated in the same way in lattice QCD, simply with different quark masses 
being chosen in the solution of the Dirac equation on a given background 
gluon field to give the quark propagator, 
and different quark propagators combined to give the current-current correlator. 

We work on ensembles (sets) of gluon field configurations that include the effect of 
$u$, $d$ and $s$ quarks in the sea with $u$ and $d$ quarks taken to have the same 
mass (denoted $m_{l}$). The configurations were generated by the 
MILC collaboration who used an $\mathcal{O}(a^2)$ improved discretisation 
of the gluon action and the improved staggered (asqtad) action for the quarks 
in the sea~\cite{Bazavov:2009bb}. 
The main parameters of the ensembles used in this work are listed 
in Table~\ref{tab:ensembles}.
The ensembles include three values of the lattice spacing, $a$, ranging 
from $a \approx$ 0.09~fm to $a \approx$ 0.045~fm. 
The sea $s$ quark mass is close to the physical $s$ quark mass 
but the $u/d$ quark masses are not at their physical values, but 
instead at the heavier $m_l/m_s$ = 0.2 point. This corresponds to a value 
for the pion mass, $M_{\pi}$, around 300 MeV. 

\begin{table}
\caption{Simulation parameters for the MILC 
$n_f=2+1$ gluon field 
ensembles that we use, labelled by set number in the first column. 
The lattice spacing values are given in units of a parameter known 
as $r_1$, derived from the static quark potential, 
in the second column~\cite{Bazavov:2009bb}. The physical value for 
$r_1$ is 0.3133(23) fm~\cite{Davies:2009tsa}. $L_s/a$ and $L_t/s$ 
give the lattice dimensions. $u_0am_l^{\sea}$ and $u_0am_s^{\sea}$ 
give the sea $u/d$ and $s$ quark masses respectively 
in lattice units in the MILC convention, 
where $u_0$ is the plaquette tadpole parameter. 
Set 1 will be referred to as `fine', set 2 as `superfine' and set 
3 as `ultrafine'. We use around 200 configurations from each ensemble 
and increase statistics by using 2 or 4 `random wall' time sources for propagators 
on each configuration. 
}  
\label{tab:ensembles}
\begin{ruledtabular}
\begin{tabular}{llllll}
Set & $r_1/a$ & $L_s/a$ & $L_t/a$ & $u_0am_l^{\text{sea}}$ & $u_0am_s^{\text{sea}}$ \\
\hline 
1 & 3.699(3) & 28 & 96  & 0.0062 & 0.031  \\
2 & 5.296(7) & 48 & 144 & 0.0036 & 0.018  \\
3 & 7.115(20) & 64 & 192 & 0.0028 & 0.014  \\
\end{tabular}
\end{ruledtabular}
\end{table}

\begin{table}
\caption{Valence masses used with the HISQ formalism 
for the quark propagators making up the pseudoscalar 
current-current correlators studied here. 
Column 1 gives the Set number (see Table~\ref{tab:ensembles}) 
and then Columns 2 and 3 give the valence $s$ and $c$ quark 
masses in lattice units respectively. Column 4 gives the list of heavy 
quark masses used and Column 5 the corresponding values (taken from Ref.~\cite{McNeile:2012qf}) for 
the mass of the pseudoscalar heavyonium meson, $\eta_h$, made from those heavy 
quark propagators.
On Set 2, we use $am_c=0.273$, which is the tuned one. The data for $am_c=0.28$ was used 
to study the dependence on the charm quark mass, which was found to be negligible (see 
Section~\ref{subsec:results}).
}  
\label{tab:masses}
\begin{ruledtabular}
\begin{tabular}{lllll}
Set & $am_s^{\text{val}}$ & $am_c^{\text{val}}$ & $am_h^{\text{val}}$ & $aM_{\eta_h}$ \\
\hline 
1 & 0.0337 & 0.413  	& 0.7 	& 1.86536(5) \\
  &  	   &   		& 0.85 	& 2.14981(5) \\
2 & 0.0228 & 0.273,0.28 & 0.564 & 1.52542(6) \\
  &  	   &  		& 0.705 & 1.80845(6) \\
  &  	   &  		& 0.85 	& 2.08753(6) \\
3 & 0.0165 & 0.195 	& 0.5  	& 1.34477(8) \\
  &  	   &  		& 0.7 	& 1.75189(7) \\
  &  	   &  		& 0.85 	& 2.04296(7) \\
\end{tabular}
\end{ruledtabular}
\end{table}

On each ensemble, we calculate quark propagators for
the $s$ quark, the $c$ quark and then for a set of heavier quark masses  
heading towards that of the $b$ quark. For these valence quarks 
we use the Highly Improved Staggered Quark (HISQ) action~\cite{hisqdef}, 
and the valence quark masses that we use are given in Table~\ref{tab:masses}. 
The tuning of the valence $s$ and $c$ quark masses to their physical 
values is discussed in Ref.~\cite{McNeile:2012qf}. 

The HISQ action 
was developed by HPQCD~\cite{hisqdef} to have very small discretisation errors and 
this has been demonstrated to be the case in many calculations 
(for example~\cite{McNeile:2010ji, fdsorig, fdsupdate}). 
This reduction in discretisation errors means more accurate results across the 
board for a given value of the lattice spacing but this is particularly 
important in the heavy quark regime. There discretisation errors are controlled 
by the quark mass in lattice units ($am$) and the HISQ action has a 
higher reach in the quark mass for a given lattice spacing than actions 
that are less improved. 
This has allowed the HPQCD collaboration to initiate a programme 
of heavy quark physics using the HISQ action that stretches from 
$c$ to $b$ physics~\cite{fdsupdate, McNeile:2010ji, bshisq}. Values 
of $ma$ that correspond to $b$ quarks can only be reached on very fine 
lattices. With results for multiple masses at multiple lattice spacings, 
however,
it is possible to fit a functional form to the results that also 
includes a functional form for discretisation effects and this 
can then be used to determine a value 
at the $b$ quark mass in the continuum limit~\cite{McNeile:2010ji, bshisq}.  

Here we are interested in correlation functions
for pseudoscalar mesons, made by multiplying together the quark propagators discussed above. 
We calculate correlators made from a variety of 
combinations of quark masses, including one heavy quark 
whose mass we denote $am_h$.  
We define the pseudoscalar current-current correlator at zero spatial momentum by 
\begin{equation}
\label{eq:corrdef}
  G(t) = a^6 \sum_{\vec{x}} (am_h)^2 \langle 0 | j_5(\vec{x}, t)  j_5(\vec{0}, 0) | 0 \rangle 
\end{equation}
where $j_5 = \overline{\psi}_1\gamma_5 \psi_2$ for the case here of two different quark flavours.  
We are working in the staggered formalism and so the $\gamma_5$ matrix is implemented 
through phase-factors~\cite{hisqdef}. Here we use the local pseudoscalar operator, 
i.e. the one with `Goldstone' taste or, using staggered spin-taste 
notation, $\gamma_5 \otimes \gamma_5$. This means that, with the mass factors 
above, the current-current correlator is absolutely normalised, and no lattice 
current renormalisation factor is needed.  

The pseudoscalar correlator defined in~(\ref{eq:corrdef}) creates a meson at 
time 0 and destroys it at time $t$. In the large $t$ limit the correlator 
is dominated by the ground-state meson with that valence quark content and spin-parity 
quantum numbers. Fitting the correlators and extracting the parameters of 
the large $t$ behaviour enables the masses and decay constants of the ground-state
pseudoscalar mesons to be determined. In Refs.~\cite{bshisq, McNeile:2012qf} HPQCD showed 
that results at multiple heavy quark masses for multiple lattice spacings 
could be combined to determine the physical dependence on the heavy quark mass of 
the heavy-strange and heavy-charm meson masses and decay constants.
This dependence could then 
be evaluated at the $b$ quark mass to yield $M_{B_s}$ and $M_{B_c}$ for a comparison 
at the few MeV level with experiment, as well as $f_{B_s}$ and $f_{B_c}$ as inputs 
to flavour physics\footnote{ 
This method has now been used 
by the Fermilab Lattice and MILC collaborations to obtain 
$f_B$ and $M_B$~\cite{Bazavov:2017lyh, Bazavov:2018omf}.}.  

Here we study the time-moments of the same 
pseudoscalar current-current correlators.
These are defined by the lattice version of Eq.~(\ref{eq:momcont}) 
\begin{align}
  G_n  &= \sum_t (t/a)^n G(t) \label{eq:def:G_n} ,
\end{align}
where $t$ respects the lattice periodicity so that~\cite{Allison:2008xk}
\begin{equation}
\label{eq:tdef}
t/a \in \{0, 1, 2 \ldots T/2a-1, 0, -T/2a+1 \ldots -2, -1\} .
\end{equation}
In contrast to the results described in the paragraph above, 
the time-moments emphasise the small-$t$ region of the correlator
where the product of the two current operators is short-distance
and an OPE is required, as described in Section~\ref{sec:theory}. 

Following Ref.~\cite{Allison:2008xk},
to reduce the discretisation errors in the time-moments we divide each moment 
by its tree level value, $G_n^{(0)}$, calculated with the 
gluon fields set to the unit matrix. For the 
heavy-charm case $G_n^{(0)}$ can be simply calculated on the 
same lattices as those used for the interacting gluon field 
ensembles. For the heavy-strange case we must use space-time 
volumes that are a factor of 3 larger in each spatial direction 
in order to eliminate finite-volume effects in the free case~\cite{Koponen:2010jy}. 
We define the reduced moment $R_n$ as
\begin{equation}
  R_n = \left(\frac{aM_{\eta_h}}{2am_{0h}}\right)^{n-4}\frac{G_n}{G_n^{(0)}} ,
  \label{eq:def:R_n}
\end{equation}
where $m_{0h}$ can either be the bare mass or the tree-level pole mass of the heavy quark, 
which differ by discretisation effects~\cite{hisqdef}. 
Which is used here is irrelevant since this factor will be cancelled below. 

The continuum limit of the lattice QCD results gives the 
value for the moments derived in continuum QCD
\begin{equation}
  R_n = \left(\frac{M_{\eta_h}}{2m_h(\mu)}\right)^{n-4} \frac{g_n(\alpha_s(\mu), \mu/m_h)}{g_n^{(0)}} .
  \label{eq:def:R_n:cont}
\end{equation}
The dimensionless function $g_n$ is the vacuum expectation value for the 
appropriate OPE for that moment and 
$g_n^{(0)}$ its tree-level component (see Eq.~(\ref{eq:hc})). 
If $g_n$ is derived in the $\overline{\text{MS}}$ scheme, then 
$m_h(\mu)$ is the heavy quark mass in that scheme. 
For the heavy-strange case, as discussed in Section~\ref{subsec:OPE}, the 
vacuum expectation value for the OPE has sizeable 
contributions from the $s$ quark condensate that we want to determine. 
In order to emphasise these, and reduce systematic effects from the heavy 
quark masses, it is useful to take a ratio of the reduced moments $R_n$ for 
heavy-strange to those of heavy-charm. 
The heavy-charm case has only a small gluon condensate contribution, 
similar to that of heavyonium discussed 
in Section~\ref{sec:theory} and so makes a useful denominator. 
Another advantage of taking this ratio is that some other systematic 
errors are partly cancelled, for example those from discretisation effects 
and missing higher orders in perturbation theory. 

For the ratio of heavy-strange moments to heavy-charm we have, in the continuum limit,
\begin{equation}
  \frac{R_n(hs)}{R_n(hc)} =  \frac{g_{n,hs}}{g_{n,hs}^{(0)}} \times \frac{g_{n,hc}^{(0)}}{g_{n,hc}}.
  \label{eq:Rnrat}
\end{equation}
As explained in Secion~\ref{subsec:OPE}, continuum QCD perturbation theory calculations 
through $\mathcal{O}(\alpha_s^2)$ as a function
of light to heavy quark mass ratio $(x)$ enable us to determine an accurate expansion of 
$g_{n,hs}$ given in Appendix~\ref{appendix:coeffs}. 
$g_{n,hs}$ consists
a leading-order (in powers of the heavy-quark mass) 
perturbative series $c_n$ with a subsidiary perturbative series $d_n$ multiplying 
the $s$ quark condensate divided by the cube of the heavy quark mass and $e_n$ 
multiplying the gluon condensate divided by the fourth power of the heavy quark 
mass (Eq.~(\ref{eq:vevope})). $g_{n,hs}^{(0)}$ is the leading 
tree-level term in $g_{n,hs}$, noting that no condensates
appear when the gluon field is set to the unit matrix.
In the heavy-charm case the same $\mathcal{O}(\alpha^2_s)$ heavy-light 
perturbation theory can be used~\cite{Hoff:2011ge, Grigo:2012ji} but now 
evaluated numerically for the specific charm
to heavy quark mass ratios 
corresponding to each set of heavy-charm moments calculated. 
The values of the coefficients in $g_{n,hc}$ for each case used here are given in 
Table~\ref{tab:g_n:hc} in Appendix~\ref{appendix:coeffs}. 
The gluon condensate contribution to the heavy-charm moments will 
be discussed further in Section~\ref{sec:analysis}. 

We will study the 
4th, 6th, 8th and 10th time-moments.
Table~\ref{tab:red_mom_ratio:lattice} gives our lattice results for the 
ratios for these moment numbers on each gluon field ensemble and 
for each heavy quark mass. Statistical uncertainties 
are shown---they are very small. In addition our fits include 
the statistical correlations between the results 
on a given ensemble; these are not included in the Table. 

\begin{table}
\newcommand{\h}{\phantom{x}}
\caption{Our lattice QCD results for the ratio of heavy-strange 
moments, $R_n$, to those of heavy-charm (columns 3, 4, 5, and 6). 
We show values for 
the ratios for the 4th, 6th, 8th, and 10th moments on each of the gluon configuration sets and for each heavy 
quark mass that we use. Uncertainties are statistical only. }  
\label{tab:red_mom_ratio:lattice}
\begin{tabular}{ll@{\quad}|llll}
 \hline\hline
         &        &     \multicolumn{3}{c}{$R_n(hs)/R_n(hc)$}    \\
\hline
    Set  & $am_h$ &  $n=4$ &   $n=6$ &   $n=8$ & $n=10$ \\
 \hline
1  &  0.7    &  1.13897(21)  & 1.16120(38)  & 1.11194(54)  &  1.03069(70) \\
   &  0.85   &  1.09811(16)  & 1.13358(30)  & 1.11783(46)  &  1.07193(62) \\
2  &  0.564  &  1.09243(42)  & 1.13547(78)  & 1.1429(11)   &  1.1230(13) \\
   &  0.705  &  1.05465(31)  & 1.09919(59)  & 1.12584(90)  &  1.1380(12) \\
   &  0.85   &  1.03024(25)  & 1.06924(45)  & 1.10044(73)  &  1.1250(10) \\
3  &  0.5    &  1.05488(41)  & 1.10513(78)  & 1.1422(12)   &  1.1619(17) \\
   &  0.7    &  1.01465(27)  & 1.05242(49)  & 1.09333(81)  &  1.1343(12) \\
   &  0.85   &  0.99954(22)  & 1.02755(37)  & 1.06201(61)  &  1.10136(91) \\
 \hline\hline
\end{tabular}
\end{table}
\begin{figure}
    \includegraphics[width=0.47\textwidth]{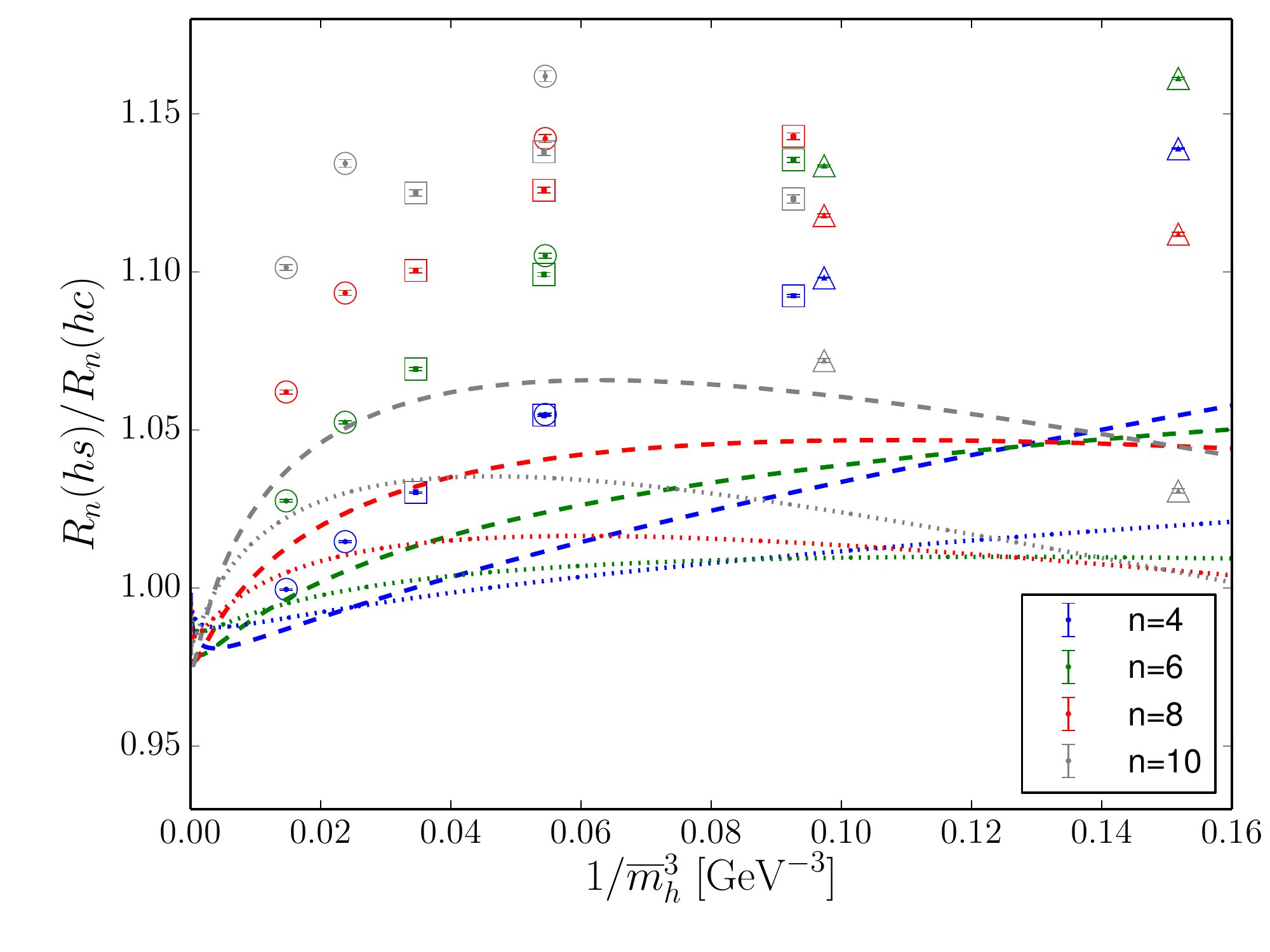}
    \caption{Ratios of reduced moments for heavy-strange to heavy-charm 
calculated on the lattice (symbols) and the ratio 
    of the appropriate perturbative series with {\it no} condensate contribution.
    The dotted and dashed lines show the ratio of the $c_n$ series for heavy-strange and 
the $g_n$ series for heavy-charm through order $\alpha_s$ and $\alpha_s^2$, respectively.
    The circles, squares, and triangles show the lattice results on ultrafine, superfine, and fine lattice ensembles respectively.
    }
   \label{fig:data-pert_nocond}
\end{figure}
Figure~\ref{fig:data-pert_nocond} shows the lattice results for $R_n(hs)/R_n(hc)$ 
at our three values of the lattice spacing.
The $x$-axis is the cube of the inverse of $\overline{m}_h \equiv {m}_h({m}_h)$, i.e. 
the heavy quark mass in the $\overline{\text{MS}}$ scheme at its own scale. 
The values for this are obtained from the values for $M_{\eta_h}$ using the function 
derived in Ref.~\cite{McNeile:2010ji}. Figure 6 of that reference shows the relationship graphically.   
$\overline{m}_h$ is proportional to $M_{\eta_h}$ (up to very small corrections that we include) over the 
range of masses we use here from ~$2m_c$ upwards.  

Also plotted on Figure~\ref{fig:data-pert_nocond} is the appropriate ratio of 
perturbative series for the ratio of moments in which {\it the condensate contributions 
are ignored}. The dotted and dashed lines show increasing orders in the perturbative 
expansion. It is clear from this that the lattice QCD results do not agree with the 
perturbation theory when the condensate contributions are ignored. At 
small $1/\overline{m}_h$ the discrepancy is linear in 
$1/\overline{m}_h^3$ also indicating that the discrepancy can be traced to a quark 
condensate contribution.  

In the next section we discuss how we can fit the lattice QCD results to the perturbative 
expansion {\it including condensate terms} derived from the OPE, allowing for 
systematic uncertainties from higher order perturbative and nonperturbative effects. 
This allows us to determine a value for the $s$ quark condensate quite accurately. It is 
clear from Figure~\ref{fig:data-pert_nocond} that this should be possible given the clear 
and significant contribution coming from that condensate.

\section{Analysis}
\label{sec:analysis}

\subsection{Fit function}
\label{subsec:fitfunction}

Our goal is to extract the strange-quark condensate from ratios of reduced moments
$R_n(hs)/R_n(hc)$ 
by comparing lattice-QCD results to theoretical expectations from the OPE 
and perturbation theory.
To this end, we need to take into account
discretization errors, scale uncertainties and mistunings of quark masses from lattice QCD 
and truncation of the OPE and truncation of 
the perturbation theory in the continuum. 
 
Combining Eqs.~(\ref{eq:Rnrat}) and~(\ref{eq:vevope}) the continuum theoretical 
expectation for $R_n(hs)/R_n(hc)$ is 
\begin{equation}
\label{eq:Rnratcont}
\frac{R_n(hs)}{R_n(hc)} = \frac{g_{n,hc}^{(0)}}{g_{n,hc}}\frac{1}{g_{n,hs}^{(0)}}\left(c_{n,hs} + d_{n,hs} \frac{\langle \overline{s}s\rangle}{\overline{m}_h^3} + e_{n,hs}\frac{\langle \alpha_s G^2 \rangle}{\overline{m}_h^4}\right)
\end{equation}
where the OPE has been truncated at $1/\mbar_h^4$ and the perturbative series for $c_n$, $d_n$ and $g_n$ 
are truncated at $\alpha_s^2$. We have changed the generic notation of $M$ for the heavy quark mass 
to the more specific $\overline{\text{MS}}$ case, where the $\overline{\text{MS}}$ mass, 
$\overline{m}_h = m_h(m_h)$ varies as our lattice 
bare heavy quark mass changes. 

The fit function for our results at non-zero lattice spacing  
then becomes
\begin{align}
\label{eq:fit-function}
  \frac{R_n(hs)}{R_n(hc)}(a, m_{\sea}) &= \frac{1}{\tilde{g}_{n,hc}} \times \nonumber \\
   &\hspace{-40pt}\left(\tilde{c}_{n,hs} + \tilde{d}_{n,hs}\frac{m_s\SBS}{x \mbar_h^4}\left[1+d_{\text{mist}}\right]  
   +\tilde{e}_{n} \frac{\langle \alpha_s G^2 \rangle}{\mbar_h^4}
   \right) \nonumber \\
  & +R_{n,\text{hoOPE}} + R_{n,\text{mist}}.
\end{align}
The output from the fit will be the fit parameter $m_s\SBS$. 
We have modified $g$, $c$, $d$ and $e$ to absorb the appropriate $g^{(0)}$ factor and 
to allow for discretisation effects in the lattice determination of $R_n(hs)$ and $R_n(hc)$ and missing 
higher order terms in the perturbation theory, to be discussed below. 
We have expressed the condensate contribution in terms of $m_s \SBS$ for later convenience, 
using the identity $m_s(\mbar_h) = x\,\mbar_h$ (since ratios of lattice quark masses in 
a particular quark formalism give the ratios of quark masses for a given continuum formalism at 
the same scale, up to 
discretisation effects~\cite{Davies:2009ih}). 
As discussed above in Section~\ref{sec:latt}, $\mbar_h$ is fixed from the $\eta_h$ mass. 
Uncertainties in the value of the lattice spacing 
(here set by $r_1$, see Table~\ref{tab:ensembles} ) are fed in at this point. 
In the third line $R_{n,\text{hoOPE}}$ models our ignorance of higher dimension condensates
and $R_{n,\text{mist}}$, along with $d_{\text{mist}}$, contain the effects of mistunings in sea and valence quark masses.
Both of these will be described further below but we begin with terms on the second line.

In Eq.~(\ref{eq:fit-function}) $\tilde{c}$ and $\tilde{d}$ are treated in a very similar way. We take
\begin{subequations}
  \begin{align}
    \label{eq:c:tilde}
    \tilde{c}_{n,hs} &= \left(\sum_{i=0}^4 \frac{c_n^{(i)}(x)}{g_{n,hs}^{(0)}(x)} \alpha_s^i(\mbar_h)\right) \times \\
    & \left[1 + \sum_{k=0}^{4} \sum_{l=0}^4 c_{n,\lat}^{(k,l)} \left(a\Lambda\right)^{2k} (am_h)^{2l} \right] , \nonumber\\
    \label{eq:d:tilde}
    \tilde{d}_{n,hs} &= \left(\sum_{i=0}^3 \frac{d_n^{(i)}(x)}{g_{n,hs}^{(0)}(x)} \alpha_s^i(\mbar_h) \right) \times \\ 
    & \left[1 + \sum_{k=0}^{4} \sum_{l=0}^4 d_{\lat}^{(k,l)} \left(a\Lambda\right)^{2k} (am_h)^{2l} \right] , \nonumber
  \end{align}
\end{subequations}
where the first term is the perturbative series for that piece of $R_n(hs)$ 
and the second term in square brackets allows for discretisation effects. 

For the perturbative series we use the expressions tabulated in Tables~\ref{tab:c_n:hs}
and~\ref{tab:d_n:hs}, for $c_{n}^{(i)}(x)$ and $d_{n}^{(i)}(x)$, respectively, with $i<3$.
$g_{n,hs}^{(0)}$ is the tree-level term in the expansion of $\mathcal{M}_n$, including the 
non-analytic terms in $x$ that become the condensate in the interacting theory; these terms 
have very little effect since $x$ is so small. 
$\alpha_s$ is evaluated in the $\overline{\text{MS}}$ scheme at the scale $\mu=\mbar_h$ 
using $n_f=3$ evolution from the starting 
point $\alpha_s(\overline{\text{MS}},n_f=3,\mu=5\,\text{GeV})=0.2034(21)$~\cite{McNeile:2010ji}. 
$x$ is the ratio of $s$ to heavy quark masses, both at the scale $\mu$. Up to discretisation 
effects this is equal to the ratio of lattice bare quark masses $x = am_s/am_h$~\cite{Davies:2009ih} 
obtained using the values 
from Table~\ref{tab:masses}.

To include higher-order terms by treating the coefficients of these terms as fit parameters,
we define, for $i\ge 3$
\begin{subequations}
  \begin{align}
    \label{eq:c:tilde:i}
    \frac{c_n^{(i)}(x)}{g_n^{(0)}(x)} & = \sum_{j=0}^{4} \tilde{c}_n^{(i,j)} (nx)^j ,\\
    \label{eq:d:tilde:i}
    \frac{d_n^{(i)}(x)}{g_n^{(0)}(x)} & = \sum_{j=0}^{1} \tilde{d}_n^{(i,j)} x^j ,
  \end{align}
\end{subequations}
with a prior of $0\pm 1$ for each $\tilde{c}_n^{(i,j)}$ with $i=3$ and $4$,
and $0\pm n^3$ for each $\tilde{d}_n^{(i,j)}$
with $i=3$. These priors are justified by comparing similar expressions for $i<3$.
Note that the expansion in Eq.~(\ref{eq:c:tilde:i}) is organized in powers of $nx$
rather than $x$ because a factor of $n$ typically accompanies each power of $x$ in
the known expansions of $c_n^{(i)}(x)$ for $i<3$. The form of expansion, however,
does not have any noticeable effects in our final result for the strange-quark condensate.

The second term, in square brackets, of each of Eqs~(\ref{eq:c:tilde}) and~(\ref{eq:d:tilde})
allows for discretisation effects in the lattice QCD calculation of $R_n(hs)$. 
We allow separately for discretisation effects 
coming from the matrix element of the unit operator and the matrix element of $\overline{\psi}\psi$ 
since they have slightly different form. We view the discretisation effects in the $d_n$ term as being 
discretisation effects in the condensate and hence we do not allow the $d_{\text{lat}}$ coefficients to 
depend on $n$. We allow for discretisation effects that depend on $a\Lambda$ and on $am_h$, taking 
$\Lambda$ = 0.3 GeV. 
Note that $c_{n, \lat}^{(0,0)}$ and $d_{\lat}^{(0,0)}$ = 0.
Since all tree-level errors are removed in the reduced moments,
we take the priors of the remaining parameters to be $\text{O}(\alpha_s)$, namely $0.0(0.3)$.

We now discuss $\tilde{g}_{n,hc}$ which allows for systematic uncertainties in 
$R_n(hc)$. 
The continuum perturbation theory for $R_n(hc)$ that we use is missing 
terms at $\alpha_s^3$ and above. However in the ratio $R_n(hs)/R_n(hc)$ those terms 
can be reabsorbed into a single perturbative series with the missing $\alpha_s^3$ terms 
from $R_n(hs)$. Hence such terms are already taken into account by allowing for 
missing higher-order perturbative terms in Eqs~(\ref{eq:c:tilde:i}) and~(\ref{eq:d:tilde:i}) 
and we do not include any additional terms for $g_{n,hc}$.  
This would also be true for discretisation effects except that the discretisation effects 
in $R_n(hc)$ can take a different form from those of $R_n(hs)$ because of the presence of 
the charm quark mass to set a higher scale for them. 
We take 
  \begin{equation}
    \label{eq:g_n:lat}
    \tilde{g}_{n,hc} =  \frac{g_{n,hc}}{g_{n,hc}^{(0)}} \times 
     \left[1 + \sum_{k=0}^{4} \sum_{l=0}^4 g_{n,\lat}^{(k,l)} (am_c)^{2k} (am_h)^{2l} \right] . 
   \end{equation}
$g_{n,hc}$ and $g_{n,hc}^{(0)}$ are evaluated from the coefficients given 
in Table~\ref{tab:g_n:hc} and the appropriate value of $\alpha_s(\mu)$.
Since we expect the leading heavy quark discretization errors to cancel 
in the ratio of $R_n(hs)$ and $R_n(hc)$
we take $g_{n,\lat}^{(0,l)}=c_{n,\lat}^{0,l}$.
As above, since all tree-level errors are removed in the reduced moments,
we take the priors of the remaining parameters to be $\text{O}(\alpha_s)$, namely $0.0(0.3)$.

In the ratio of heavy-strange moments to heavy-charm, both numerator and denominator
have contributions from the gluon condensate. 
For the heavy-strange case, we have determined the leading order coefficient, $e_n$, 
which is $1/(12\pi)$, independent of $n$. 
The heavy-charm case is similar to that of heavyonium, discussed in Section~\ref{sec:theory}
in that the contribution comes from gluon propagators where $k^2 \rightarrow 0$. 
Taking the results of Ref.~\cite{Reinders:1980wk} for the vacuum polarisation function 
for a valence quark and antiquark of different masses 
and expanding in powers of $z=q^2/m_h^2$ allows us to determine the gluon condensate contribution to 
the heavy-charm moments. For simplicity we also make an expansion in powers of $1/m_c$ and collect 
terms in $x_c^{-1}$ and $x_c^{0}$ where $x_c = m_c/m_h$.    
We collect all contributions
to the gluon condensate in one term with coefficient $\tilde{e}_n$ (see Eq.~(\ref{eq:fit-function})). 
Then
\begin{equation}
  \label{eq:tilde-e}
  \tilde{e}_{n} = \frac{1}{12\pi}  \left( \frac{1}{g_{n,hs}^{(0)}} 
  - \frac{\tilde{c}_{n,hs}}{g_{n,hc}} \left(\frac{m_h}{m_c} - \frac{n-2}{4} \right)\right) .
\end{equation}

In order to treat $m_s\SBS$ and $\langle \alpha_s G^2\rangle$ as fit parameters in Eq.~(\ref{eq:fit-function}) 
we must fix their scales, rather than allowing them to run with $\mu$.
To this end, we first rewrite them as
\begin{subequations}
  \begin{align}
    \label{eq:PBP:mu2nu}
    m_s \SBS(\mu) &= 
    m_s \SBS(\nu) + \zeta_1(\mu,\nu) m_s^4(\mu) , \\
    \label{eq:aGG:mu2nu}
    \langle \alpha_s G^2\rangle (\mu) &=
    \langle \alpha_s G^2\rangle (\nu) \\
    &+ \zeta'_{1}(\mu,\nu) m_s^4(\mu) + \zeta'_{2}(\mu,\nu) \times m_s \SBS(\nu) . \nonumber
  \end{align}
\end{subequations}
Exploiting the perturbative expansion in Eqs.~(\ref{eq:PBP:pert})
and~(\ref{eq:gluecondensate}), we obtain
\begin{subequations}
  \begin{align}
    \label{eq:zeta1_mu_nu}
    \zeta_1(\mu,\nu) &= \frac{L}{4\pi^2} \bigg[ 6 + \frac{\alpha_s(\mu)}{\pi} (8 + 24L)   \nonumber \\
    &\hspace{-20pt}+ \left(\frac{\alpha_s(\mu)}{\pi} \right)^2
      \bigg( 33.275 - \frac{29}{2}n_f + (155 - \frac{14}{3}n_f)L \nonumber \\
    &+ (108 - \frac{8}{3}n_f)L^2\bigg) 
    \bigg] , \\
    \label{eq:zeta1_prime_mu_nu}
    \zeta'_1(\mu,\nu) &= - \frac{2n_m  L}{\pi} \left(\frac{\alpha_s(\mu)}{\pi} \right)^2 (1 - 3 L ) ,\\
    \label{eq:zeta2_prime_mu_nu}
    \zeta'_2(\mu,\nu) &= - 8 n_m \pi L \left(\frac{\alpha_s(\mu)}{\pi} \right)^2 ,
  \end{align}
\end{subequations}
where $L = \ln(\mu/\nu)$. The effect of the mixing of $m_s \SBS$ and $\langle \alpha_s G^2 \rangle$
with the identity or with each other, 
which generates the running in Eqs.~(\ref{eq:PBP:mu2nu}) and~(\ref{eq:aGG:mu2nu}), 
is very small since powers of the $s$ quark mass appear. Any mixing of $\langle \alpha_s G^2 \rangle$ 
with the sea $u/d$ quark condensate is completely negligible as a result. 
Then, setting $\mu=\mbar_h$ and $\nu=2$~GeV,
we plug Eqs.~(\ref{eq:PBP:mu2nu}) and~(\ref{eq:aGG:mu2nu}) 
in Eq.~(\ref{eq:fit-function}).
Finally, we treat $m_s\SBS(2~\GeV)$ as a fit parameter with a prior of $0\pm1~\GeV^4$.
For the gluon condensate we set the prior of $\langle\frac{\alpha_s}{\pi}G^2\rangle (2~\GeV)$
to $0\pm0.012~\GeV^4$~\cite{Ioffe:2005ym}.

To allow for the effect of condensates of higher dimension than the ones covered by our OPE, 
we use
\begin{equation}
 R_{n,\text{hoOPE}}  = \sum_{k=5}^{10} r^{(k)}_{n,OPE} \left(\frac{n \Lambda}{\mbar_h}\right)^{k} .
\end{equation}
In the fit we set $\Lambda=0.3~\GeV$ and take $0\pm1$ for the prior values
of the dimensionless coefficients $r^{(k)}_{n,OPE}$. 
Note that we include a factor of $n$ for each power of $\Lambda$ because we see
such a pattern for some fermionic condensates (see, for example,~\cite{Jamin:1992se}). 
Our fit is insensitive to how many additional condensates we include beyond $k=8$. 
Our results do not give a signal for any condensate other than the quark condensate. 

Note that, as discussed in Section~\ref{subsec:OPE} we can neglect the impact of a light 
quark condensate from the sea $u/d$ quarks. This can appear at $\alpha_s^2$ but 
only in combination with the $u/d$ 
quark mass and then divided by four powers of $\mbar_h$. Any such term would have
negligible impact here.   

For effects of mistuning in the sea-quark masses 
in Eq.~(\ref{eq:fit-function}) we use
\begin{equation}
 \label{eq:R_mist}
 R_{n,\text{mist}} =
  \sum_{k=1}^2 r_{n,\text{mist}}^{(k)} \left(\frac{\delta m_\text{sea}}{m_{s,\text{phys}}}\right)^k ,
\end{equation}
where
\begin{equation}
  \delta m_\text{sea}
   = \left(2 m_{l, \text{sea}} + m_{s, \text{sea}}\right)
   - \left(2 m_{l, \text{phys}} + m_{s, \text{phys}}\right) .
\end{equation}
The values of $\delta m_\text{sea}/m_{s,\text{phys}}$ for the ensembles
that we use are tabulated in Table~VI of Ref.~\cite{fdsupdate}.
Since we are using unphysial sea $u/d$ quark masses the values of 
$\delta m_\text{sea}$ are non-zero and vary from 0.31 to 0.59. 
We also use, following Ref.~\cite{McNeile:2012xh},
a prior of $0.00(1)$ for $r_{n,\text{mist}}^{(k)}$.
Eq.~(\ref{eq:R_mist}) incorporates the leading mistuning effects into our analysis.
Since the strange-quark condensate itself can be affected by mistuning in the sea
quark masses as well the valence strange-quark mass, we allow for this through 
the parameter $d_{\text{mist}}$ in Eq.~(\ref{eq:fit-function}), with
\begin{align}
 \label{eq:d_mist}
 d_{\text{mist}} = 
    d_{\text{mist}}^{(1)} \left(\frac{\delta m_\text{sea}}{m_{s,\text{phys}}}\right) 
    + d_{\text{mist}}^{(2)} \left(\frac{M^2_{\eta_s} - M^2_{\eta_{s,\text{phys}}}}{1~\GeV^2}\right) .
\end{align}
In this analysis we use $M_{\eta_{s,\text{phys}}} = 0.6858$~GeV~\cite{Davies:2009tsa},
and following Ref.~\cite{McNeile:2012xh}
we use a prior of $0.0(1)$ for $d_{n,\text{mist}}^{(1)}$ and $0.0(5)$ for $d_{\text{mist}}^{(2)}$.

Using the fit function of Eq.~(\ref{eq:fit-function}) 
we proceed to fit 
the lattice QCD results of Table~\ref{tab:red_mom_ratio:lattice} to obtain a fitted 
value for $m_s\SBS$(2 GeV).  

\subsection{Results}
\label{subsec:results}
\begin{figure}
   \includegraphics[width=0.47\textwidth]{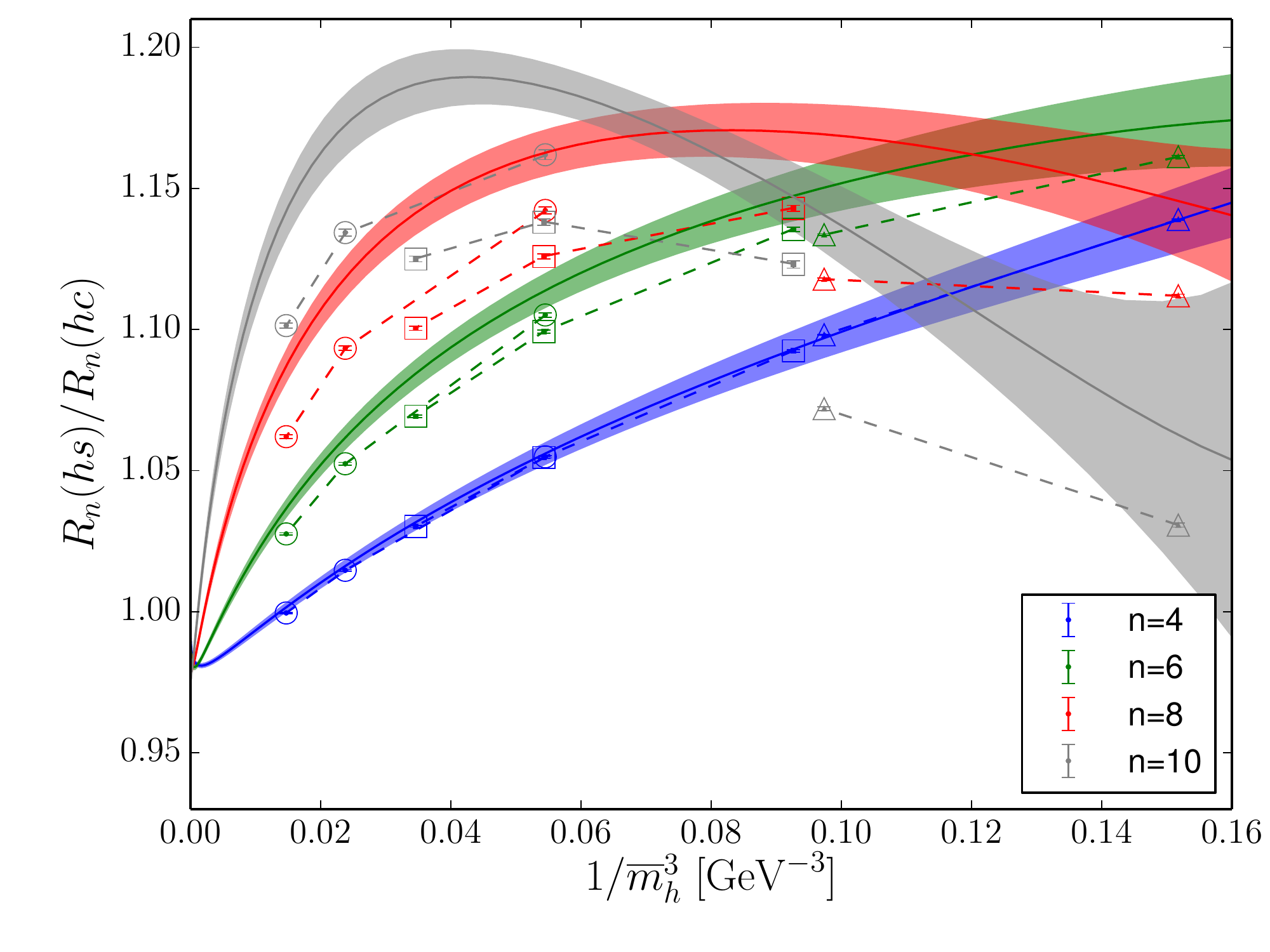}
   \caption{The lattice data for ratios of reduced moments of heavy-strange and heavy-charm
      correlators at 3 lattice spacings (symbols), and the continuum extrapolation from our fit (coloured bands).
The circles, squares and triangles show the lattice results on the ultrafine, superfine and fine 
ensembles respectively. 
     The dashed lines join the fitted values at each data point. 
     }
   \label{fig:hs-hc_extrap}
\end{figure}

As described on Sec.~\ref{sec:latt} we have three sets of lattice gluon field configurations
for which we have calculated $R_n(hs)/R_n(hc)$ 
for $n=$4, 6, 8 and 10; results are tabulated on Table~\ref{tab:red_mom_ratio:lattice}.
To fit these we use the fit function described in Section~\ref{subsec:fitfunction} and perform a combined fit
as well as separate fits to the data for each $n$ value.
Figure~\ref{fig:hs-hc_extrap} shows the lattice values and
the band gives the fit results, extrapolated to zero lattice spacing and physical sea 
masses, from the combined fit. The combined fit has 
$\chi^2/\text{dof} = 7/32$. If we drop the quark condensate from the fit function 
the combined fit gives  
$\chi^2/\text{dof} = 102/32$ showing our inability to fit the lattice results without including 
this term. The fit yields a value for the fit parameter $m_s\SBS$(2 GeV).

\begin{figure}
    \includegraphics[width=0.48\textwidth]{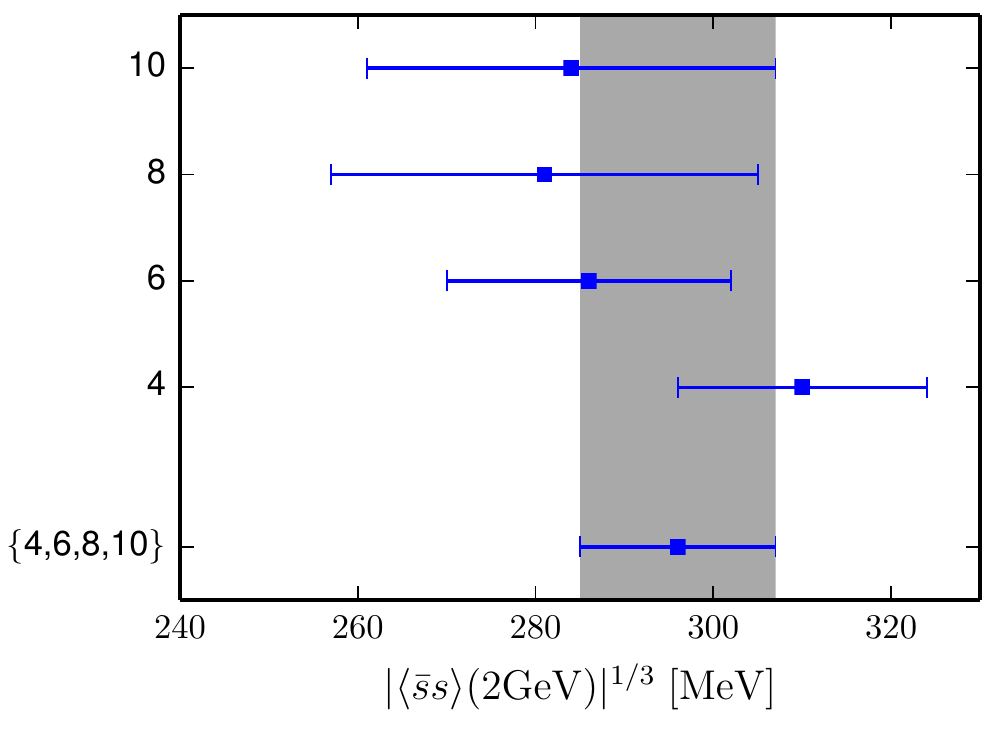}
    \caption{The stability of the cube root of $\SBS(2\GeV)$ as we change the lattice QCD results used in our work.
    Here `$\{4,6,8,10\}$' denotes the result from a combined fit to the ratio of moments for $n$ in $\{4,6,8,10\}$.
    The gray band corresponds to the combined fit.
    }
   \label{fig:stability}
\end{figure}
\begin{table}
\caption{Sources of uncertainty for the strange-quark condensate, $\SBS^{1/3}$, determined 
from $R_n(hs)/R_n(hc)$.
The uncertainties are given as a percentage of the final value.}
\label{tab:error-budget}
  \begin{tabular}{l@{\quad}@{\quad}l}
    \hline\hline
     \multicolumn{2}{c}{percentage uncertainty in $|\SBS|^{1/3}$} \\
    \hline
    lattice QCD results			&       0.86 \\
    $a \rightarrow 0$ extrapolation 	&       1.66 \\
    higher order pert. th.    		&       1.48 \\
    gluon condensate    		&       1.41 \\
    higher order condensates    	&  	1.75 \\
    mistunings  			&       1.11 \\
    $m_s(2\GeV)$      			&       0.47 \\
    $r_1$ uncertainty			&	0.73 \\
    $r_1/a$ uncertainties		&	1.02 \\
    $\alpha_s$ uncertainty		&	0.23 \\
    \hline
      Total				&       3.72 \\
    \hline\hline
  \end{tabular}
\end{table}

To obtain the strange-quark condensate we divide $m_s\SBS(2~\GeV)$
by $m_s(2~\GeV;~n_f=3) = 92.2(1.3)~\MeV$~\cite{McNeile:2010ji},
which is determined using the lattice-QCD ensembles used in this analysis
and additional ensembles. We ignore the correlation between 
the value of the strange-quark mass and our final result
for $m_s\SBS(2~\GeV)$ because the uncertainty in
$m_s$ is considerably smaller than that in $m_s\SBS(2~\GeV)$.
Our value for the strange-quark condensate from 
a combined fit to $R_n(hs)/R_n(hc)$ with $n=$4, 6, 8 and 10 is then  
\begin{equation}
  \label{eq:SBS:from_combined_hs/hc}
  \SBS(2 \GeV) = -(296(11)\MeV)^3 .
\end{equation}

Figure~\ref{fig:stability} shows the stability of $\SBS(2\GeV)$ as we change the lattice data used in our work.
It shows the results when a simultaneous fit or separate fits to the 4th, 6th, 8th and 10th time-moments are performed.
The uncertainties in the separate fits are larger for $n=$ 8 and 10 than for 
$n=4$ and 6.

The dominant sources of error for $\SBS(2\GeV)$, obtained from 
the simultaneous fits to the 4th to 10th time-moments,
are tabulated in Table.~\ref{tab:error-budget}.
The contributions from most sources 
are of similar size, about 1 to 2 percent.
The dominant four uncertainties come, not surprisingly, from the gluon condensate, higher 
order condensates, higher order terms in perturbation theory and discretisation 
effects ($a \rightarrow 0$ extrapolation). 

To investigate the stability of our analysis we performed the following tests:
\begin{itemize}
\item %
to investigate the impact of $\alpha_s^3$ corrections, we fixed $\tilde{c}_n^{(3,0)}$
in Eq.~\eqref{eq:c:tilde:i} to the known $\alpha_s^3$ correction to $R_n(hs)$
for massless light quarks~\cite{Maier:2015jxa}, ignoring the corresponding 
(unknown) correction in $R_n(hc)$.
We find a $1\sigma$ shift in $\SBS$ under this test,
but it should be emphasized that the $\alpha_s^3$ correction being added 
here is incomplete, and therefore not correct. Indeed we would expect some 
cancellation of $\alpha_s^3$ terms between $R_n(hs)$ and $R_n(hc)$ as 
we see at $\mathcal{O}(\alpha_s)$ and $\mathcal{O}(\alpha_s^2)$ (see 
Tables~\ref{tab:c_n:hs} and~\ref{tab:g_n:hc}). 
\item %
to test the impact of varying the charm quark mass, we repeated the analysis 
using the alternative `ultrafine' data
with the mistuned mass value $am_c=0.28$ instead of those with the tuned 
$am_c=0.273$ (see Table~\ref{tab:masses}).
The value of $\SBS$ decreases by $0.2\sigma$.
\item %
to reduce the effect of lattice artifacts, we dropped the `fine' data. 
The value of $\SBS$ decreased by $0.6\sigma$.
\item %
to test sensitivity to the presence of the gluon condensate, 
we repeated the analysis doubling the prior width of the gluon condensate.
The value of $\SBS$ then increases by $0.4\sigma$.
\end{itemize}
These tests show that our result for the strange-quark condensate is stable 
under variations in the fit function and lattice data.
It is also noteworthy that the posterior values of almost all fit parameters 
are within the $1\sigma$ width of their prior distributions.
This implies that there is no tension between our choice of priors and the lattice data.

\section{Conclusions}
\label{sec:conclusions}

\begin{figure}
    \includegraphics[width=0.47\textwidth]{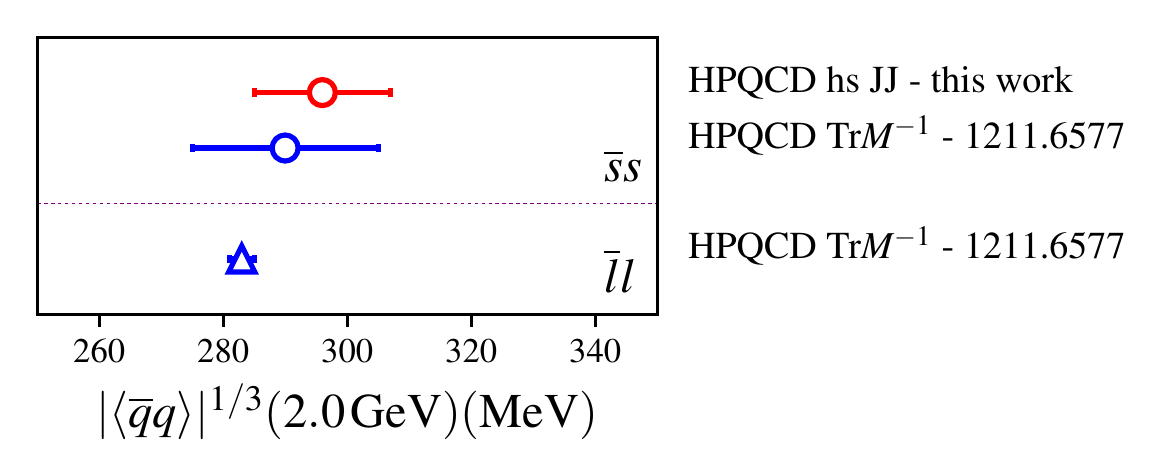}
    \caption{Comparison of the new result for $\SBS$ obtained here (red circle) with the earlier result 
from HPQCD using the $\Tr M^{-1}$ (blue circle)~\cite{McNeile:2012xh}. We also show the result for the light 
quark condensate (blue triangle) from that work. All results are in the $\overline{\text{MS}}$ scheme at 
a scale of 2 GeV. Our new results have $n_f=3$; the results in~\cite{McNeile:2012xh} have $n_f=4$.    
    }
   \label{fig:condcomp}
\end{figure}

We have developed a new method for determining the light quark condensate 
using time-moments of heavy-light current-current correlators calculated in 
lattice QCD. We fit lattice QCD results for the heavy-strange case 
as a function of heavy quark mass and lattice spacing to the 
expectation based on an OPE accurate through $\alpha_s^2$ and 
$1/\mbar_h^4$. 
We are able to extract a result for $\SBS$ with a total uncertainty of 11\%, 
or 3.7\% in $|\SBS|^{1/3}$. Our result, in the $\overline{\text{MS}}$ scheme at 
2 GeV and with $n_f=3$ is 
\begin{equation}
\label{eq:result}
  \SBS(2 \GeV) = -(296(11)\MeV)^3 .
\end{equation}
The error budget is given in Table~\ref{tab:error-budget}. 

The reasons that this accuracy is possible in this calculation are:
\begin{itemize}
\item the light quark condensate is the leading (in powers of $1/\mbar_h$) condensate in 
the heavy-light moments. It appears at tree-level, suppressed by three powers of the 
heavy quark mass but with no suppression by 
the light quark mass.
\item  we have an $\alpha_s^2$-accurate OPE so that the coefficients of both the leading `perturbative' 
term (from the unit operator) and the coefficient of the quark condensate are under good 
perturbative control and the coefficient of the gluon condensate, at the next order in $1/\mbar_h$,
is known to leading order. 
\item we have accurate lattice QCD results for the time-moments at multiple values of 
the lattice spacing and multiple heavy 
quark masses. The HISQ action that we use has small discretisation errors (although they are visible
here) and allows us to push to high quark masses. Using multiple heavy quark masses allows the identification 
of the quark condensate term from its functional dependence. 
Indeed we have shown that it is not possible 
to fit the results without including a quark condensate (Section~\ref{subsec:results}). 
\item  we have used a ratio of heavy-strange to heavy-charm correlator moments that removes systematic 
uncertainties from overall powers of the heavy quark mass that would otherwise appear. We also fit 
multiple moments simultaneously to improve the accuracy.  
\end{itemize}

We can compare our result for the strange quark condensate to that obtained 
earlier by HPQCD from a completely independent method~\cite{McNeile:2012xh}. 
That method used lattice QCD results for the vacuum expectation value of the 
trace of the quark propagator ($\Tr(M^{-1})$) along with an 
$\mathcal{O}(\alpha_s)$-accurate OPE for that case.
In the OPE for the $\Tr(M^{-1})$ the relative behaviours of the unit operator 
and $\overline{\psi}\psi$ terms are rather different from the method introduced 
here. Here the short-distance scale is physical because it is set by $\mbar_h$ and we 
can use lattice results at a variety of lattice spacing values and masses to pin down 
the condensate term. In the $\Tr(M^{-1})$ method the 
short-distance scale is set by the lattice spacing.  
Then the term coming from the unit operator is suppressed, 
relative to $\PBP$, by a power of the light quark mass but it diverges, relatively, 
by two powers of the inverse lattice spacing as $a\rightarrow 0$. 
This means that the most useful results are those on coarse lattices using 
improved actions for small discretisation effects and this is the approach taken 
in~\cite{McNeile:2012xh}. 
Results with multiple light quark masses were used to fit/constrain 
higher order terms but the strange quark condensate was less accurately determined than 
that of the light $u/d$ quarks, because of the size of the contribution from the unit 
operator. 

The value for the strange quark condensate obtained from the $\Tr(M^{-1})$ method
is -$(290(15)\,\text{MeV})^3$~\cite{McNeile:2012xh}. Our new result in Eq.~(\ref{eq:result}) agrees well 
with this and is more accurate.  
Note that the earlier result includes 4 flavours of sea quarks and here we include 
3 flavours. We expect the impact of that change to be negligible, however, given that 
it produces a 0.2\% change in $m_s$ from perturbation theory. 
This agreement between two very different 
calculations is strong validation of the OPE 
approach for short-distance quantities in fully nonperturbative QCD. 
The analysis here also `closes the loop' on understanding both the small-$t$ and large-$t$ 
behaviour of the heavy-light current-current correlators from lattice QCD, as discussed in Section~\ref{sec:latt} 
and as has already been achieved for heavyonium correlators~\cite{Allison:2008xk}. 

Figure~\ref{fig:condcomp} compares our new result for $|\SBS|^{1/3}$ to the earlier 
result from Ref.~\cite{McNeile:2012xh} 
for both $|\SBS|^{1/3}$ and $|\langle \overline{l}l\rangle|^{1/3}$, where $l$ has the 
average of the $u$ and $d$ quark masses. Our new 
results confirm that $\SBS$ and $\langle \overline{l}l\rangle$ are close in value, with 
a 1$\sigma$ preference for $\SBS$ to be slightly larger in magnitude. 
Given that the chiral condensate 
(at zero quark mass) is smaller in magnitude than that 
of the light quark~\cite{McNeile:2012xh}, it is clear that the magnitude of the condensate increases 
with quark mass for small quark mass. 
At heavy quark mass, however, the condensate magnitude must fall with quark mass~\cite{Generalis:1983hb}.  
Where, in the middle of this picture, the strange quark condensate sits is not yet 
completely clear. 
We believe that our new method can be used in future to explore/constrain further the dependence of the 
condensate on quark mass between light and strange.

\acknowledgments

We are grateful to the MILC collaboration for the use of 
their configurations and to Matthias Steinhauser for useful discussions. 
The computing for this project used the Argonne Leadership Computing Facility at the 
Argonne National Laboratory and the Data Analytic system at the University of Cambridge, 
operated by the High Performance Computing Service on behalf of the STFC DiRAC 
HPC Facility. It was funded by a BIS national E-infrastructure capital grant 
and STFC capital and operations grants to DiRAC.
We are grateful to the HPCS support staff for assistance. 
Funding for this work came from the 
Science and Technology Facilities Council.

%\clearpage
\appendix
\section{Coefficients of the OPE for heavy-light current-current correlators}
\label{appendix:coeffs}

\begin{table}
\caption{The coefficients $c_n^{(i)}$ of the moments of heavy-light current-current correlators, 
defined in Eqs~(\ref{eq:OPE1}) and~(\ref{eq:cn}).
Here $n_l$, $n_m$ and $n_h$ are the number of massless, mass $m$ and mass $M$ sea quarks.
}
\label{tab:c_n:hs}
 \begin{tabular}{|c|c|l|}
 \hline\hline
 $n$ & $i$ &  $~c^{(i)}_n(x)$ \\
\hline
 \multirow{6}{*}{4}
 & 0		 & $0.01267+0.01267 x-0.03800 x^2+0.25330 x^3-0.46861 x^4$ \\
 & 1	 & $0.02989+0.09955 x-0.23253 x^2+0.57550 x^3-1.01067 x^4$ \\
 & 2  & $\left(0.00605 n_h+0.00197 n_l+0.00197 n_m+0.03733\right)$ \\
 &	         & $+x \left(0.00581 n_h-0.03152 n_l-0.03152 n_m+1.05365\right)$ \\
 &		 & $+x^2 \left(-0.01456 n_h+0.06909 n_l+0.08353 n_m-2.27380\right)$ \\
 &		 & $+x^3 \left(0.09401 n_h+0.42679 n_l-0.40543 n_m+5.54802\right)$ \\
 &		 & $+x^4 \left(-0.18395 n_h-0.90926 n_l+0.59586 n_m-9.19481\right)$ \\
 \hline
 \multirow{6}{*}{6}
 & 0		 & $0.00317+0.00633 x-0.02533 x^2+0.29763 x^3-0.68708 x^4$ \\
 & 1	 & $0.00853+0.04872 x-0.15004 x^2+0.31603 x^3-0.71574 x^4$ \\
 & 2	 & $\left(0.00043 n_h+0.00092 n_l+0.00092 n_m+0.03585\right)$ \\
 &		 & $+x \left(0.00103 n_h-0.01694 n_l-0.01694 n_m+0.60059\right)$ \\
 &		 & $+x^2 \left(-0.00336 n_h+0.04330 n_l+0.04621 n_m-1.68628\right)$ \\
 &		 & $+x^3 \left(0.05508 n_h+0.78889 n_l-0.16592 n_m+4.20811\right)$ \\
 &		 & $+x^4 \left(-0.14984 n_h-1.98801 n_l+0.17277 n_m-8.21087\right)$ \\
 \hline
 \multirow{6}{*}{8}
 & 0		 & $0.00127+0.00380 x-0.01900 x^2+0.33183 x^3-0.91949 x^4$ \\
 & 1	 & $0.00200+0.02529 x-0.09410 x^2-0.16782 x^3+0.36671 x^4$ \\
 & 2  & $\left(-0.00019 n_h+0.00116 n_l+0.00116 n_m+0.01080\right)$ \\
 &		 & $+x \left(-0.00036 n_h-0.00896 n_l-0.00896 n_m+0.35061\right)$ \\
 &		 & $+x^2 \left(0.00183 n_h+0.02274 n_l+0.02170 n_m-1.19699\right)$ \\
 &		 & $+x^3 \left(-0.00446 n_h+1.27450 n_l+0.22966 n_m+1.69697\right)$ \\
 &		 & $+x^4 \left(-0.02649 n_h-3.72325 n_l-0.88075 n_m-2.99808\right)$ \\
 \hline
 \multirow{6}{*}{10}
 & 0		 & $0.00063+0.00253 x-0.01520 x^2+0.35969 x^3-1.16393 x^4$  \\
 & 1	 & $-0.00004+0.01313 x-0.05450 x^2-0.82509 x^3+2.38624 x^4$ \\
 & 2  & $\left(-0.00026 n_h+0.00117 n_l+0.00117 n_m+0.00101\right)$ \\ 
 &		 & $+x \left(-0.00084 n_h-0.00437 n_l-0.00437 n_m+0.21065\right)$ \\
 &		 & $+x^2 \left(0.00479 n_h+0.00653 n_l+0.00389 n_m-0.83194\right)$ \\
 &		 & $+x^3 \left(-0.07786 n_h+1.86152 n_l+0.74592 n_m-1.12449\right)$ \\
 &		 & $+x^4 \left(0.19427 n_h-6.21161 n_l-2.66365 n_m+5.57420\right)$ \\
\hline\hline
 \end{tabular}
\end{table}

\begin{table*}
\caption{The $d_n$ coefficient of the quark condensate in the OPE for 
moments of heavy-light current-current correlators. $d_n$ is defined in 
Eq.~(\ref{eq:OPE1}) and the individual pieces by analogy to that 
for the $c_n^{(i)}$ in Eq.~(\ref{eq:cn}). 
}
\label{tab:d_n:hs}
 \begin{tabular}{|c|c|l|}
 \hline\hline
 $n$ & $i$ &  $~d^{(i)}_n(x)$ \\
 \hline
 \multirow{3}{*}{4}
 & 0	&  $  -1  + 2 x $  \\
 & 1	&  $  4   - 7.667 x$  \\
 & 2	&  $ (18.890 - 0.387n_h - 4.754n_l - 2.754n_m) + x (-38.881 + 0.803n_h + 9.415n_l + 6.290n_m)$ \\
 \hline
 \multirow{3}{*}{6}
 & 0	&  $  -1  + 2.5 x $  \\
 & 1	&  $ 6.5  - 15.333 x $  \\
 & 2	&  $(25.479 - 0.178n_h - 6.506n_l - 4.506n_m) + x (-65.770 + 0.513n_h + 15.969n_l + 12.159n_m)$ \\
 \hline
 \multirow{3}{*}{8}
 & 0	&  $  -1     + 3 x  $  \\
 & 1	&  $ 9.067 - 25.5 x$  \\
 & 2	&  $ (26.692 + 0.030 n_h - 8.347n_l - 6.347n_m) + x (-85.46557 + 0.023 n_h + 24.433 n_l + 19.883 n_m)$  \\
 \hline
 \multirow{3}{*}{10}
 & 0	&  $  -1  +3.5 x  $  \\
 & 1	&  $ 11.667  -38.183 x $  \\
 & 2	&  $ (21.798 + 0.237n_h - 10.241n_l - 8.241n_m) + x (-88.697 - 0.672n_h + 34.821n_l + 29.498n_m)$ \\
 \hline\hline
 \end{tabular}
\end{table*}

\begin{table*}
\caption{The coefficients $g^{(i)}_n$ for the time-moments of 
the heavy-charm current-current correlators used in this analysis. 
The values given in brackets in the column headings correspond 
to the mass ratios for charm to heavy. }
\label{tab:g_n:hc}
 \begin{tabular}{|c|c|l@{\quad}l@{\quad}l@{\quad}l@{\quad}l@{\quad}l@{\quad}l@{\quad}l|}
 \hline\hline
 $n$ & $i$ 	&  $g_n(0.229)$  & $g_n(0.486)$ & $g_n(0.390)$ &	$g_n(0.484)$ &	$g_n(0.321)$ &	$g_n(0.590)$ &	$g_n(0.279)$ &	$g_n(0.387)$ \\
 \hline
 \multirow{3}{*}{4}
 & 0 &  0.013  & 0.0107 & 0.0116 & 0.0107 & 0.0123 & 0.00962 & 0.0127 & 0.0117 \\
 & 1 &  0.0343 & 0.0261 & 0.0292 & 0.0261 & 0.0315 & 0.0231  & 0.0328 & 0.0293 \\
 & 2 &  0.0736 & 0.0307 & 0.0445 & 0.031  & 0.0563 & 0.0193  & 0.0642 & 0.0449 \\
 \hline
 \multirow{3}{*}{6}
 & 0 &  0.00305 & 0.00187 & 0.00228 & 0.00188 & 0.00261 & 0.0015 & 0.00281 & 0.00229 \\
 & 1 &  0.00906 & 0.00564 & 0.00673 & 0.00566 & 0.00766 & 0.0047 & 0.00829 & 0.00677 \\
 & 2 &  0.0443  & 0.0235  & 0.0295  & 0.0236  & 0.0351  & 0.0186 & 0.039   & 0.0297 \\
 \hline
 \multirow{3}{*}{8}
 & 0 &  0.00107 & 0.000475 & 0.000654 & 0.000478 & 0.000816 & 0.000335 & 0.000928 & 0.00066 \\
 & 1 &  0.00193 & 0.00101  & 0.00125  & 0.00101  & 0.00149  & 0.000815 & 0.00167  & 0.00126 \\
 & 2 &  0.014   & 0.00636  & 0.00836  & 0.00639  & 0.0103   & 0.00481  & 0.0118   & 0.00844 \\
 \hline
 \multirow{3}{*}{10}
 & 0 &  0.000449 & 0.000142 & 0.000221 & 0.000143 & 0.000303 & 8.77e-05 & 0.000364 & 0.000224 \\
 & 1 &  1.68e-05 & 9.27e-05 & 6.47e-05 & 9.24e-05 & 3.52e-05 & 0.000101 & 2e-05    & 6.35e-05 \\
 & 2 &  0.00383  & 0.00154  & 0.00211  & 0.00155  & 0.00268  & 0.00111  & 0.00313  & 0.00213 \\
 \hline\hline
 \end{tabular}
\end{table*}

The OPE leads to an expansion for the time-moments of heavy-light current-current 
correlators that can be organized as (repeating Eq.~(\ref{eq:vevope}))
\begin{eqnarray}
\label{eq:OPE1}
 \frac{M^{n-4}}{n!}{\mathcal{M}_n} &=&  g_n(x, \alpha_s(\mu), \mu/M) \\ 
&=& c_n + d_n \frac{\PBP}{M^3} 
  +  e_n \frac{{\langle \alpha_s G^2 \rangle}}{M^4}
  + \mathcal{O}\left(x^5\right) . \nonumber
\end{eqnarray}
The short-distance coefficients $c_n$, $d_n$, $e_n$ and their higher order counterparts
are power series in $\alpha_s$ with only polynomial dependence on $x$.
So, for example, 
\begin{equation}
\label{eq:cn}
c_n = c_n^{(0)} + c_n^{(1)} \asPI + c_n^{(2)} \left(\asPI\right)^2 + \ldots
\end{equation}
with $c_n^{(i)}$ a polynomial in $x$. We work in the $\overline{\text{MS}}$ scheme 
at a scale $\mu$, and will use $\mu=M(\mu)\equiv M$ so that there are no logarithms of $\mu/M$ in
the coefficients. The logarithms can be reconstructed by evolving $\alpha_s$ and $M(\mu)$.   
Since we work to $\mathcal{O}(x^4)$ in the original series, the $c_n^{(i)}$ have terms 
up to and including $x^4$ and the $d_n^{(i)}$ up to and including $x$.  
$\PBP$ and $\langle \alpha_s G^2 \rangle$ are evaluated at scale $\mu$ in 
Eq.~(\ref{eq:OPE1}). 

We use $n_l$, $n_m$, and $n_h$ to denote the number of massless, mass $m$ and heavy mass $M$ quarks in the sea, respectively.
In the case of heavy-strange correlators calculated on gauge 
configurations with $(2+1)$-flavors that we use here (see Section~\ref{sec:latt}), 
we set $n_l=2$, $n_m=1$, and $n_h=0$. 

Table~\ref{tab:c_n:hs} shows the expressions for the $c_n^{(i)}$ 
and Table~\ref{tab:d_n:hs} for the $d_n^{(i)}$, derived from the 
perturbative expressions in Refs.~\cite{Hoff:2011ge, Grigo:2012ji, Kneur:2015dda}. 
With an $\mathcal{O}(\alpha^2_s)$ analysis we can only 
access $e_n^{(0)}$ and we find $e_n^{(0)} = 1/(12\pi)$, for all $n$.  

Note that in Eqs.~(\ref{eq:PBP:pert}) and~(\ref{eq:gluecondensate}) the gauge coupling $\alpha_s$
corresponds to a theory with $n_l+n_m$ active quarks, \textit{i.e.}, $n_h=0$, while
the gauge coupling in Ref.~\cite{Grigo:2012ji} corresponds to a theory with $n_l+n_m+n_h$ active
quarks. Therefore, when the matching is performed at scale $\mu$, one should use~\cite{Rodrigo:1995az}
\begin{eqnarray}
  \alpha_s(\mu; n_l+n_m) &=& \alpha_s(\mu; n_l+n_m+n_h) \times \\
&&\hspace{-60pt}\left( 1 + \frac{n_h}{3} \alpha_s(\mu; n_l+n_m+n_h)\ln(M/\mu)
  + \mathcal{O}(\alpha^2_s) \right) \nonumber
   \label{eq:couple:n_h}
\end{eqnarray}
to match the two couplings.

In part of our analysis in Section~\ref{sec:latt} we use 
heavy-charm current-current correlator moments. The perturbative expansion 
for these is given by 
\begin{eqnarray}
\label{eq:hc}
 \frac{M^{n-4}}{n!}{\mathcal{M}_n} &=&  g_n(x, \alpha_s(\mu), \mu/M) \\
  &=& g_n^{(0)} + g_n^{(1)}\asPI + g_n^{(2)}\left(\asPI\right)^2 + \ldots \nonumber 
\end{eqnarray}
where we ignore, for now, condensate contributions. That issue 
is considered as part of the systematic error analysis in Section~\ref{sec:analysis}. 
We use the perturbative expansion given 
in Refs.~\cite{Hoff:2011ge, Grigo:2012ji} for unequal mass quarks, evaluating each of 
the coefficients at the mass ratio corresponding to that of the lattice charm and 
heavy quark masses. The values of the $g^{(i)}_n$ at each of these ratios are 
tabulated in Table~\ref{tab:g_n:hc}. Note that here, for 2+1 light flavours in the sea, 
$n_l=3$, $n_m=0$ and $n_h=0$.

%\clearpage
\bibliographystyle{apsrev4-1}
\bibliography{paper_ref}

\end{document}